\documentclass[10pt]{iopart}

\usepackage{booktabs}
\usepackage{xy}

\usepackage{amsmath}
\usepackage{amsthm}
\usepackage{amssymb}

\input xy
\xyoption{all}

\usepackage{hyperref}  

\numberwithin{equation}{section}
\newtheorem{theorem}{Theorem}
\newtheorem{proposition}{Proposition}
\newtheorem{definition}{Definition}

\DeclareMathOperator{\spanvettoriale}{span}

\begin{document}

\title[Combinatorics of transformation coefficients in Brauer algebras]{Combinatorics of transformations from standard to non-standard bases in Brauer algebras}

\author{Vincenzo Chilla}

\address{Dipartimento di Fisica ``Enrico Fermi'', Universit\`a di Pisa and Sezione INFN - Largo Bruno Pontecorvo 3, 56127 Pisa, Italy}
\ead{chilla@df.unipi.it}

\begin{abstract}
Transformation coefficients between {\it standard} bases for irreducible representations of the Brauer centralizer algebra $\mathfrak{B}_f(x)$ and {\it split} bases adapted to the $\mathfrak{B}_{f_1} (x) \times \mathfrak{B}_{f_2} (x) \subset \mathfrak{B}_f (x)$ subalgebra ($f_1 +f_2 = f$) are considered. After providing the suitable combinatorial background, based on the definition of \emph{$i$-coupling} relation on nodes of the \emph{subduction grid}, we introduce a generalized version of the \emph{subduction graph} which extends the one given in \emph{J. Phys. A: Math. Gen.} $\mathbf{39}$ $7657-7668$ for symmetric groups. Thus, we can describe the structure of the subduction system arising from the linear method and give an outline of the form of the solution space. An ordering relation on the grid is also given and then, as in the case of symmetric groups, the choices of the phases and of the free factors governing the multiplicity separations are discussed.  \\   
\phantom{a} 
\end{abstract}


\section{Motivations}
Racah-Wigner calculus for classical Lie groups (unitary, orthogonal and symplectic groups) plays a fundamental role in many areas of physics and chemistry. Orthogonal and symplectic Racah-Wigner calculus arises, for example, in the description of symmetrized orbitals in quantum chemistry and in fermion and boson many-body theory~\cite{fukutome}, grand unification theories for elementary particles~\cite{gellmann}, supergravity~\cite{gellmann1}, interacting boson and fermion dynamical symmetry models for nuclei~\cite{arima,navratil}, nuclear symplectic models~\cite{rosensteel, castanos}, and so on. 
\par 
In particular, Racah coefficients and other recoupling coefficients of unitary $SU(n)$, orthogonal $SO(n)$ and symplectic $Sp(2m)$ groups of different rank are quite useful when calculating energy levels and transition rates in atomic, molecular and nuclear theory (for example, in connection with the Jahn-Teller effect and structural analysis of atomic shells, see Judd and co-workers~\cite{ju, judd2} and, for a description of multi-bosonic and multi-fermionic systems and applications in the microscopic nuclear theory, consider~\cite{vana, het}), and in conformal field theory~\cite{petkova}. 
\par 
There are many approaches to the Racah coefficients, but the problem is that there is not a general method for treating various kind of coupling and recoupling issues. Any given technique applies only to a particular problem and for a particular group. Not only the tecniques for dealing with unitary, orthogonal and symplectic groups all drastically differ from each other, but the methods for finding the various Wigner coefficients also vary from one to the other. Furthermore, analytical expressions are difficult to come by for the general Lie groups, mainly because there is a \emph{multiplicity problem} in the reduction of Kronecker products of pair of irreducible representations. Some missing labels need to be added in, for which a procedure is often difficult to do systematically. Finally, although several efficient computer codes and numerical procedures exist, they often do not permit any insight in the mathematical structure of such coefficients and, however, we still need a general and efficient closed algorithm.  
\par 
The goal to provide a systematic and comprehensive approach to deal with the structure of coupling and recoupling coefficients for classical Lie groups is not been achieved. However, the most promising strategy for this purpose seems to be the one building on the well-known and tight connection between symmetric and unitary groups which is called in literature \emph{Schur-Weyl duality} and which was first pointed out by Schur in the beginning of the twentieth century~\cite{schur}. This observation was ten years later developed by Brauer~\cite{brauer} who found the full centralizer algebra for orthogonal and symplectic groups and gave the construction of the full centralizer algebras for the classical series of the Lie groups.
\par 
Kramer~\cite{kramer} used explicit transformations between the bases defined in terms of different symmetric group chains (so called \emph{Gelfand-Tzetlin chains}) to define his $f$ \emph{symbol} (our subduction factor) for a symmetric group. He showed that the such symbols were equivalent to recoupling coefficients ($6j$ and $9j$ symbols) for any unitary group and furthermore that $f$ symbols were also equal to coupling coefficients for $U(p+q) \supset U(p) \times U(q)$. Later~\cite{pandai} these results were generalized to Brauer centralizer algebras and to the corresponding ortho-symplectic groups, making the problem of finding coupling and recoupling coefficients for classical Lie groups \emph{equivalent} to the \emph{subduction problem} for centralizer algebras.  
\par 
In this paper, we choose an algebraic approach to the subduction problem in Brauer algebras $\mathfrak{B}_{f}(x) \downarrow \mathfrak{B}_{f_1}(x) \times \mathfrak{B}_{f_2}(x)$ ($f_1 + f_2 =f$) and we provide a combinatorial description of the equation system arisen from the linear equation method~\cite{pandai1}. By solving the subduction problem for such centralizer algebras, one has the way for a unified approach to the coupling and recoupling issue in classical Lie groups. 
\par  
Following the layout of~\cite{chilla}, in section 2, we give the irreducible representation of Brauer algebras and, by introducing the concept of \emph{permutation lattice}, we present the explicit action of the generators on the invariant irreducible modules. In section 3, we provide the explicit form for the subduction equations and, in section 4, we link them to the concept of a \emph{subduction graph} which generalizes the one given in~\cite{chilla}. By using the subduction graph approach, in section 5 we are able to describe the structure of the solution space for the subduction problem. We recognize that the subduction space can be built on four tipical configurations in the $i$-layer: the \emph{crossing}, the \emph{horizontal} and \emph{vertical bridges} and the \emph{singlet}. In section 6, we discuss the general orthonormalized form for the subduction coefficients and we define a suitable ordering relation on permutation lattices and on the \emph{grid} (and thus on the set of the subduction coefficients) which is necessary to fix the choice of the phases (for istance, the \emph{Young-Yamanouchi phase convention}) and of the free factors governing the multiplicity separations. Finally, in section 7, some main perspectives are briefly discussed.   

\section{Irreducible representations of Brauer algebras}
The Brauer algebra $\mathfrak{B}_{f}(n)$ is algebraically defined by the mean of $2f-2$ generators $\{g_{1}, \ g_{2},\ldots,\ g_{f-1}, \ e_{1}, e_{2},\ldots,e_{f-1}\}$  satisfying the following relations~\cite{Pan}:
\begin{align}
g_{i}g_{i+1}g_{i} & = g_{i+1}g_{i}g_{i+1} \\ 
g_{i} g_{j}       & = g_{j} g_{i}   \ \ \ \ \ \ \ \ \ \ \ \  \textmd{with $|i-j| \geq 2$} \\
e_i g_i           & = e_i \\
e_i g_{i-1} e_i   & = e_i \\
e_i^2             & = x e_i \\
g_i^2             & = 1 .
\end{align} 
In an equivalent way, $\mathfrak{B}_{f}(n)$ can be defined as the $\mathbb{C}(x)$-span of the \emph{f-diagrams}~\cite{wenzl}. We remark that the first $f-1$ generators $g_i$ also generate the subalgebra $\mathbb{C}\mathfrak{S}_f \subset \mathfrak{B}_f(x)$ (i.e. the group algebra associated to the symmetric group $\mathfrak{S}_f$).
\par 
As pointed out in~\cite{Pan}, it is known that $\mathfrak{B}_f(x)$ is semisimple, i.e. it is a direct sum of full matrix algebras over $\mathbb{C}$, when $x$ is not an integer or is an integer with $x \geq f-1$, otherwise $\mathfrak{B}_f(x)$ is not semisimple. Whenever $\mathfrak{B}(x)$ is semisimple, its irreducible representation can be labelled by a Young diagram with $f$, $f-2$, $f-4$, $\dots$, 1 or 0 boxes. It can be seen that by removing the generators $e_{f-1}$ and $g_{f-1}$, $\{g_1, \ g_2, \ldots, g_{f-2}, \ e_{1}, \ e_2 \ldots, e_{f-2} \}$ generate $\mathfrak{B}_{f-1}(x)$. By doing so repeatedly, one can establish the standard Gelfand-Tzetlin chain $\mathfrak{B}_f(x) \subset \mathfrak{B}_{f-1}(x) \subset \ldots \subset \mathfrak{B}_2(x)$. It defines the {\it standard basis} of $\mathfrak{B}_f(x)$. Let $\Upsilon_f$ be the set of all Young diagrams with $k\leq f$ boxes such that $k\geq 0$ and $f-k$ is even. If $\mathfrak{B}_f(x)$ is semisimple, it decomposes into a direct sum of full matrix algebras $\mathfrak{B}_{f,\lambda}(x)$, where $\lambda \in \Upsilon_f$. If $[f, \lambda]$ is a simple $\mathfrak{B}_{f,\lambda}(x)$ module, it decomposes as a $\mathfrak{B}_{f-1,\lambda}(x)$ in to a direct sum 
\begin{equation}
[f,\lambda] = \bigoplus_{\mu \leftrightarrow \lambda} \ [f-1,\mu]
\label{br}
\end{equation}
where $[f-1,\mu]$ is a simple $\mathfrak{B}_{f-1,\mu}(x)$ module and $\mu$ runs through all diagrams obtained by removing or (if $\lambda$ contains less than $f$ boses) adding a box to $\lambda$.
\par 
In what follows, we always assume that $\mathfrak{B}_f(x)$ is semisimple.
\subsection{Generalized tableaux}
The branching rule given in (\ref{br}) allow us to label the elements of the standard basis for an irreducible representation (irrep) $[f,\lambda]$ of the Brauer algebra $\mathfrak{B}_f(x)$ by defining a {\it generalized} Young tableau which is associated to the concept of {\it Bratteli diagram}~\cite{leducram}. 
\par 
A Bratteli diagram $A$ is a graph with vertices from a collection of sets $\check{A}_k, \ k\geq 0$, and edges that connect vertices in $\check{A}_k$ to vertices in $\check{A}_{k+1}$. One assumes that the set $\check{A}_0$ contains a unique vertex denoted by $\emptyset$. It is possible that there are multiple edges connecting any two vertices. We shall call the vertices {\it shapes}. The set $\check{A}_k$ is the set of shapes on $level \ k$. If $ \lambda \in \check{A}_k$ is connected by an edge to a shape $\mu \in \check{A}_{k+1}$ we usually write $\lambda \leq \mu$. 
\par 
A {\it multiplicity free Bratteli diagram} is a Bratteli diagram such that there is at most one edge connecting any two vertices. Here, we {\it assume} that all Bratteli diagrams are multiplicity free. In fact, the Bratteli diagrams, which are more interesting for our purposes, are multiplicity free and arise naturally in the representation theory of centralizer algebras. 
\begin{figure}
\begin{center}
$
\xymatrix@C=1pc{
0 & &  &  & *{\emptyset} \ar@{->}[d]                       & & &  \\    
1 & &  &  & *{[1]} \ar@{->}[dl] \ar@{->}[dr] & & &  \\
2 & &  & *{[2]} \ar@{->}[dl] \ar@{->}[dr] &   & [1,1] \ar@{->}[dl]  \ar@{->}[dr] & & \\
3 & & *{[3]} \ar@{->}[d] \ar@{->}[dr] &  & *{[2,1]} \ar@{->}[dl] \ar@{->}[d] \ar@{->}[dr] & & *{[1,1,1]} \ar@{->}[dl] \ar@{->}[d] & \\
4 &  & *{[4]} \ar@{->}[dl] \ar@{->}[d] & *{[3,1]} \ar@{->}[dl] \ar@{->}[d] \ar@{->}[dr]  &  *{[2,2]} \ar@{->}[dl]
\ar@{->}[dr] & *{[2,1,1]} \ar@{->}[dl] \ar@{->}[d] \ar@{->}[dr]  & *{[1,1,1,1]} \ar@{->}[d] \ar@{->}[dr] & \\
5 & *{[5]} & *{[4,1]} & *{[3,2]} & *{[3,1,1]} & *{[2,2,1]} & *{[2,1,1,1]} & *{[1,1,1,1,1]}
}
$
\end{center}
\caption{\small{First five levels of the Bratteli diagram describing the branching rule of $\mathbb{C} \mathfrak{S}_f$ centralizer algebras. $\check{A}_k$ is the set of the partitions of $k$. So, the shapes are Young diagrams and $\lambda \in \check{A}_k$ is connected to $\mu \in \check{A}_{k+1}$ by an edge if $\mu$ can be obtained from $\lambda$ by adding one box.}}
\label{brattsn}
\end{figure}
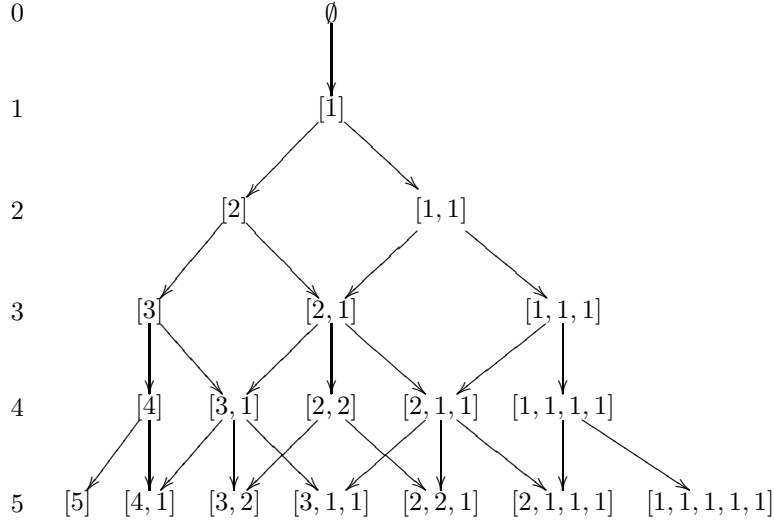
In figure \ref{brattsn} we show the Bratteli diagram describing the branching rule for the Gelfand-Tzetlin chain of $\mathbb{C} \mathfrak{S}_f$ centralizer algebras.
\par
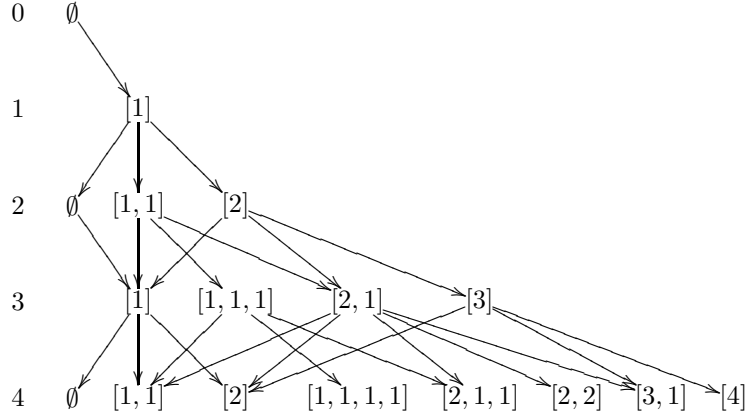
\begin{figure}
\begin{center}
$
\xymatrix@C=1pc{
0 & *{\emptyset} \ar@{->}[dr] & & & & & & & \\
1 & & *{[1]} \ar@{->}[dl] \ar@{->}[d] \ar@{->}[dr] & & & & & & \\
2 &*{\emptyset} \ar@{->}[dr] & *{[1,1]} \ar@{->}[d] \ar@{->}[dr] \ar@{->}[drr]&*{[2]} \ar@{->}[dl] \ar@{->}[dr] \ar@{->}[drr] &  & & & & \\
3 & & *{[1]} \ar@{->}[dl] \ar@{->}[d] \ar@{->}[dr] & *{[1,1,1]} \ar@{->}[dl] \ar@{->}[dr] \ar@{->}[drr]& *{[2,1]} \ar@{->}[dll] \ar@{->}[dl] \ar@{->}[dr] \ar@{->}[drr] \ar@{->}[drrr] & *{[3]} \ar@{->}[dll] \ar@{->}[drr] \ar@{->}[drrr] &  & & & \\
4 & *{\emptyset}& *{[1,1]} &*{[2]} &*{[1,1,1,1]} &*{[2,1,1]} & *{[2,2]}& *{[3,1]} & *{[4]}
}
$
\end{center}
\caption{\small{First four levels of the Bratteli diagram describing the branching rule of $\mathfrak{B}_f(x)$ centralizer algebras. Here, the shapes are Young diagrams such that $\lambda \in \check{A}_k$ is connected to $\mu \in \check{A}_{k+1}$ by an edge if $\mu$ can be obtained from $\lambda$ by adding or deleting one box.}}
\label{brattbr}
\end{figure}

 
Let $A$ be a multiplicity free Bratteli diagram and let $\lambda \in \check{A}_k$ and $\mu \in \check{A}_l$ where $k<l$. A {\it path} from $\lambda$ to $\mu$ is e sequence of shapes $\lambda^{(i)}$, $k \leq i  \leq l$, $P=(\lambda^{(k)}, \lambda^{(k+1)}, \ldots, \lambda^{(l)})$ such that $\lambda= \lambda^{(k)} \leq \lambda^{(k+1)}, \ldots, \lambda^{(l)}=\mu $ and $\lambda^{(i)} \in \check{A}_i$. A generalized \emph{tableau} $\tau$ of shape (diagram) $\lambda$ is a path from $\emptyset$ to $\lambda$, $\sigma=(\lambda^{(0)}, \lambda^{(1)}, \ldots, \lambda^{(k)})$, such that $\emptyset= \lambda^{(0)} \leq \lambda^{(1)}, \ldots, \lambda^{(k-1)} \leq \lambda^{(k)}=\lambda$ and $\lambda^{(i)} \in \check{A}_i$ for each $1 \leq i \leq l$. The branching rule for Brauer algebras given in the previous subsection can also be described by a suitable multiplicity free Bratteli diagram, as shown in figure \ref{brattbr}.
\subsection{Permutation lattices}
Let $\mathcal{W}$ be the set of all finite {\it words} composed of elements of $\mathbb{Z}\setminus\{0\}$. We define a {\it counting} function on $\mathcal{W}$ as follows:
\begin{equation}
\hat{\#}_{w}(k) = \#_{w}(k) - \#_{w}(-k), 
\end{equation}
where $\#_{w}(k)$ represents the number of times that $k\in \mathbb{Z}\setminus\{0\}$ appears in the word $w$.
\par 
Observe that, if $w$ is the empty word $\emptyset$, $\#_{w}(k)$ vanishes by definition for all  $k\in \mathbb{Z}\setminus\{0\}$. Denoting by $w^{(i)}$ the word obtained from $w$ only considering the first $i$ elements and neglecting the other ones (note that if $w$ is a word composed of $f$ elements then $w^{(f)}=w$), we give the following definition:
\begin{definition}
A \emph{permutation lattice} of order $f$ is a word $w$ composed of $f$ elements such that
\begin{equation}
\hat{\#}_{w^{(i)}}(1) \geq \hat{\#}_{w^{(i)}}(2) \geq \hat{\#}_{w^{(i)}}(3) \geq \ldots \geq 0
\label{perlat}
\end{equation}
for all $1 \leq i \leq f$. The tuple $\lambda=[\hat{\#}_{w}(1), \hat{\#}_{w}(2), \hat{\#}_{w}(3), \ldots, \hat{\#}_{w}(l)]$, where $\hat{\#}_{w}(l)$ is the last element different from zero in the sequence (\ref{perlat}), is called \emph{shape} or \emph{diagram} of $w$.
\end{definition}
For istance, the word $w=(1,1,2,-1,1,-2,2)$ is a permutation lattice of order $7$ with diagram $[2,1]$, but $v=(1,2,1,-1,2,1,3)$ is not a permutation lattice because $\hat{\#}_{v^{(5)}}(1) \ngeq \hat{\#}_{v^{(5)}}(2)$.  
\par 
We observe that if $w$ has positive elements only, the previous definition of permutation lattice becomes the usual one given, for example, in~\cite{Fulton}.
\subsection{Labelling for the Gelfand-Tzetlin base} 
For each tableau $\tau=(\lambda^{(0)}, \lambda^{(1)}, \lambda^{(2)}, \ldots,\lambda^{(f-1)}, \lambda^{(f)})$, where  $\lambda^{(0)}=\emptyset$ and $\lambda^{(f)}=\lambda$ being the shape, we can associate the $f$-tuple (or word) $w(\tau)=(w_1, w_2, \ldots, w_f)$ as follows: 
\begin{equation}
w_k=\left\{
\begin{array}{cc}
h  	& \textmd{if the Young diagram $\lambda^{(k+1)}$ is obtained from $\lambda^{(k)}$} \\
   	& \textmd{by adding one box to the $h$th row (from the top of the diagram)} \\
-h 	& \textmd{if the Young diagram $\lambda^{(k+1)}$ is obtained from $\lambda^{(k)}$} \\
	& \textmd{by deleting one box from the $h$th row (from the top of the diagarm)}
\end{array}
\right. .
\label{word}
\end{equation}
\par 
Building on the previous definitions, the following proposition easily holds:
\begin{proposition}
$\tau$ is a tableau of the Bratteli diagram for the Brauer algebra $\mathfrak{B}_f(x)$ (see figure \ref{brattbr}) if and only if $w(\tau)$ is a permuation lattice of order $f$. Furthemore, the diagram of $\tau$ coincides with the diagram of $w(\tau)$.  
\end{proposition}
Therefore, permutation lattices provide a labelling scheme for the irreducible representations of Brauer algebras. In fact, given the irrep $[f,\lambda]$ of $\mathfrak{B}_f(x)$, the relative Gelfand-Tzetlin base vectors can be labelled by all permutation lattices of order $f$ and diagram $\lambda$ (denoted by $\lambda_f$ if we need to specify the level $f$ in the Bratteli diagram).  
\par 
The dimensions of irreps of $\mathfrak{B}_f(x)$, $[f,\lambda]$, can be computed by using Bratteli diagrams inductively. One can prove that the dimension formula can be espressed~\cite{Pan} as  
\begin{equation}
\dim(\mathfrak{B}_f(x); [f,\lambda]) = \frac{f!}{(f-2k)! (2k)!!} \dim(\mathfrak{S}_{f-2k};[\lambda]) 
\label{dimbr}
\end{equation}
where $f-2k$ is the number of boxes which compose the diagram $\lambda$ and $\dim(\mathfrak{S}_{f-2k};[\lambda])$ is the dimension for the irrep $[\lambda]$ of $\mathfrak{S}_{f-2k}$ which can be further be espressed, for example, by Littlewood-Robinson formula for irreps of symmetric groups.
\par 
It should be noted that (\ref{dimbr}) provides the number of permutation lattices of order $f$ and diagram $\lambda$ once we know the number of Standard Young tableaux with diagram $\lambda$. Furthermore, we remark that the labelling scheme and the decomposition for $\mathfrak{B}_f(x)$ are the same of the corresponding quantum deformation algebras (i.e. Birman-Wenzl algebras) if the quantum deformation parameters $q$ and $r$ are not roots of unity. Thus (\ref{dimbr}) also applies to Birman-Wenzl algebras when $q$ and $r$ are not roots of unity.
\subsection{Transpose permutation lattice}
It is easily seen that the following proposition holds for any permutation lattice $w=(w_1, w_2, \ldots, w_f)$ of order $f$.
\begin{proposition}
The word $\bar{w}=(\bar{w}_1, \bar{w}_2, \dots, \bar{w}_f )$ defined by
\begin{equation}
\bar{w}_i = \hat{\#}_{w^{(i-1)}}(w_i) + \theta(w_i),
\end{equation}
where 
\begin{equation}
\theta(w_i)=
\left\{
\begin{array}{cc}
1 & \textmd{if $w_i > 0 $} \\
0 & \textmd{if $w_i < 0$}
\end{array}
\right.
\end{equation}
is a permutation lattice of order $f$ (note that $w^{(0)}$ is the empty word $\emptyset$ and $\hat{\#}_{w^{(0)}}(w_i)=0$ for all $0\leq i \leq f$).
\end{proposition}
We call $\bar{w}$ the {\it transpose permutation lattice} of $w$ and denote it by $w^{t}$. One may show the following desired involution property
\begin{equation}
(w^{t})^{t} = w.
\label{invtras}
\end{equation}
Relation (\ref{invtras}) generalizes the corresponding one for a standard Young tableau written as permutation lattice.
\subsection{Some combinatoric functions for permutation lattices}
Following~\cite{leducram} and rewriting in our ``permutation lattice language'', we define
\begin{equation}
\bigtriangledown_i(w) = (w^{t}_i - w_i -x) + x \theta(w_i)
\end{equation}
where $w=(w_1, w_2, \ldots, w_f)$ is, as usual, a permutation lattice of order $f$ and $1 \leq i \leq f$. Here $x\in\mathbb{C}$ is a parameter (the same defining $\mathfrak{B}_f(x)$). 
\par 
Given two permutation lattices of order $f$, $u$ and $v$, with the same diagram $\lambda$, we can construct the ``diamond'' function as follows:
\begin{equation}
\diamondsuit_i(u,v) = \bigtriangledown_{i+1}(u) - \bigtriangledown_i(v).
\label{diamond}
\end{equation}
We note that, if $u_h = v_h$ for all $h\neq i$ and $h\neq i+1$, the following simmetry property holds:
\begin{equation}
\diamondsuit_i(u,v) = \diamondsuit_i(v,u).
\end{equation}
Furthermore, the diamond function is related to the usual axial distance for standard Young tableaux. Precisely, given a tableau $\sigma$ and the associated permutation lattice $w$, we have that
\begin{equation}
d_i(w)=\diamondsuit_i(w,w),
\label{dass}
\end{equation}
where $d_i(w)$ denotes the axial distance between the boxes $i$ and $i+1$ in the Young diagram of $\sigma$. So, the diamond function provides a way to extend the definition of axial distance to permutation lattices. In fact, the axial distance between $i$ and $j$ in the permutation lattice $w$ can be defined by
\begin{equation}
d_{ij}(w) =
\left\{
\begin{array}{cc}
\sum_{h=i}^{j-1} \ \diamondsuit_i(w,w)   & \textmd{if $i<j$}\\
0				         & \textmd{if $i=j$}\\
- \sum_{h=j}^{i-1} \ \diamondsuit_i(w,w) & \textmd{if $i>j$}	
\end{array}
\label{}  
\right.
\end{equation}
\par 
Finally, following~\cite{king}, for each Young diagram $\lambda$, one can define the polinomials
\begin{equation}
P_{\lambda}(x)= \prod_{(i,j)\in \lambda} \frac{x-1+d(i,j)}{h(i,j)},
\label{pol}
\end{equation}
where $h(i,j)$ is the the ``hook'' function evaluated for the box in the $i$th row and $j$th column of $\lambda$: 
\begin{equation}
h(i,j)=\lambda_i + \lambda'_j -i -j + 1
\end{equation}
and $d(i,j)$ is given by
\begin{equation}
d(i,j)=
\left\{
\begin{array}{cc}
\lambda_i + \lambda_j -i -j + 1 & \textmd{if $i \leq j$} \\
- \lambda'_i - \lambda'_j + i + j - 1 & \textmd{if $i > j$}
\end{array}
\right.
\end{equation}
with $\lambda_i$ denoting the length of the $i$th row and $\lambda'_j$ the length of the $j$th column in $\lambda$. 
\par 
Note that the polinomial function (\ref{pol}) has the property that $P_{\lambda}(2n+1)$ is the dimension of each irreducible representation $V^{\lambda}$ of the orthogonal group $SO(2n+1)$.
\subsection{Explicit actions}
Now we can give the explicit action~\cite{leducram} for the generators of Brauer algebras $\mathfrak{B}_f(x)$ on the Gelfand-Tzetlin basis parameterized by permutation, but first we need the following definitions:
\begin{definition} 
Let $u=(u_1, u_2, \ldots, u_f)$ and $v=(v_1, v_2, \ldots, v_f)$ be two permutation lattices of order $f$ which have the same diagram $\lambda$. We say that $u$ is \emph{$i$-coupled} to $v$ (or that $u$ and $v$ are \emph{$i$-coupled}) if 
\begin{equation}
u_h = v_h  
\end{equation}
for all $h \in \{1, \ldots, i-1, i+2, \ldots, f\}$ and we denote such a relation by $u \stackrel{i}\leftrightarrow v$.
\end{definition}
\begin{definition}
Let $u=(u_1, u_2, \ldots, u_f)$ and $v=(v_1, v_2, \ldots, v_f)$ be two $i$-coupled permutation lattices. We say that $u$ is \emph{$\bar{i}$-coupled} to $v$ (or that $u$ and $v$ are \emph{$\bar{i}$-coupled}) if 
\begin{equation}
u_i=-u_{i+1}, \ v_i=-v_{i+1} 
\end{equation}
and we denote such a relation by $u \stackrel{\bar{i}}\leftrightarrow v$.
\end{definition}
Finally, it will be useful to introduce the \emph{$(i\setminus\bar{i})$-coupling} as follow: given two pemutation lattices $u$ and $v$, we say that $u$ is $(i\setminus\bar{i})$-coupled to $v$ (or that $u$ and $w$ are \emph{$(i\setminus\bar{i})$-coupled}) if it results $u \stackrel{i}\leftrightarrow v$ but \emph{not} $u \stackrel{\bar{i}}\leftrightarrow v$. We denote such a relation by $u \stackrel{i\setminus\bar{i}}\leftrightarrow v$.
\par 
Of course the previous definitions can also be given for tableaux. We simply say that two tableau $\sigma$ and $\tau$ are $i$-coupled or $\bar{i}$-coupled if the corresponding permutation lattices $w(\sigma)$ and $w(\tau)$ are $i$-coupled or $\bar{i}$-coupled, respectively. Note that the $i$-coupling relation just given coincides with the classical $i$-coupling relation given in~\cite{chilla} if $\sigma$ and $\tau$ are standard Young tableaux.  
\par 
Let $[f,\lambda]$ be an irrep for the Brauer algebras $\mathfrak{B}_f(x)$. The standard Gelfand-Tzetlin base for such an irrep can be parameterized by all permutation lattices $w$ of order $f$ and diagram $\lambda$: $\{|f; \lambda;w\rangle\}$. The explicit action of the $\mathfrak{B}_f(x)$ generators $g_i$ and $e_i$ on such vectors is described by the following theorem~\cite{leducram}:
\begin{theorem}
Let $u$ and $v$ two permutation lattices of order $f$ and diagram $\lambda$ and $|f; \lambda; u \rangle, \ |f; \lambda; v \rangle$ two standard base vectors for the irrep $[f,\lambda]$ of $\mathfrak{B}_f(x)$. 
\begin{itemize}
\item If $u$ and $v$ are \emph{not} $i$-coupled, then
\begin{equation}
\langle f; \lambda; u| g_i |f; \lambda; v \rangle = \langle f; \lambda; u| e_i |f; \lambda; v \rangle = 0.
\label{a1}
\end{equation}
\item If $u$ and $v$ are $i$-coupled but \emph{not} $\bar{i}$-coupled, then
\begin{equation}
\langle f; \lambda; u | g_i  |f; \lambda; v \rangle =
\left\{
\begin{array}{cc}
\frac{1}{d_i(u)} & \textmd{if $u = v$} \\
\sqrt{1-\frac{1}{d_i^{2}(u)}} & \textmd{if $u \neq v$}
\end{array}
\right. 
\label{a2}
\end{equation}
and 
\begin{equation}
\langle f;\lambda; u | e_i  |f;\lambda; v \rangle = 0,
\label{a3}
\end{equation}
where $d_i(u)=\diamondsuit_i(u,u)$ (as in (\ref{dass})).
\item If $u$ and $v$ are $\bar{i}$-coupled, then
\begin{equation}
\langle f; \lambda; u | g_i  |f;\lambda; v \rangle =
\left\{
\begin{array}{cc}
\frac{1}{\diamondsuit_i(u,u)} (1-\frac{P_{Y(u^{(i)})}(x)}{P_{Y(u^{(i-1)})}(x)}) & \textmd{if $u=v$} \\
-\frac{1}{\diamondsuit_i(u,v)} \frac{\sqrt{P_{Y(u^{(i)})}(x) P_{Y(v^{(i)})}(x)}}{P_{Y(u^{(i-1)})}(x)} & \textmd{if $u\neq v$}
\end{array}
\right.
\label{a4}
\end{equation}
and 
\begin{equation}
\langle f; \lambda; u | e_i  |f; \lambda; v \rangle = \frac{\sqrt{P_{Y(u^{(i)})}(x) P_{Y(v^{(i)})}(x)}}{P_{Y(u^{(i-1)})}(x)},
\label{a5}
\end{equation}
where $Y(w)$ denotes the diagram of the permutation lattice $w$.
\end{itemize}
\label{action}
\end{theorem}
We observe that the previous theorem provides the same action for $g_i$ given in~\cite{chilla} if $u$ and $v$ are not $\bar{i}$-coupled (as the case of standard Young tableaux). Furthermore, we can easily verify that both $g_i$ and $e_i$ are \emph{hermitian} operator on the invariant irreducible modules of Brauer algebras.


\section{The subduction problem}
Subdction coefficients (SDCs) for the reduction $[f,\lambda] \downarrow \mathfrak{B}_{f_1}(x) \times \mathfrak{B}_{f_2}(x)$ ($f_1 + f_2 = f$) define the base changing matrix which makes explicit the decomposition in block-diagonal form:
\begin{equation}
[f,\lambda]=\bigoplus_{\lambda_1,\lambda_2} \{f_1,f_2;\lambda; \lambda_1, \lambda_2\} \ [f_1,\lambda_1]\otimes[f_2,\lambda_2].
\end{equation}
Therefore, each non-standard base vector for $[f,\lambda]$ is given by the tensor product of two standard base vectors for the irreps $[f_1,\lambda_1]$ and $[f_2,\lambda_2]$. $\{f_1,f_2;\lambda; \lambda_1, \lambda_2\}$ denotes the Clebsch-Gordan series which provide the multiplicity of $[f_1,\lambda_1]\otimes[f_2,\lambda_2]$ in $[f,\lambda]$.
\par 
The irreps of $\mathfrak{B}_{f_1}(x) \times \mathfrak{B}_{f_2}(x)$ may be labelled by $(f_1, f_2; \lambda_1, \lambda_2)$ with $\lambda_1$ and $\lambda_2$ suitable partitions (shapes). In the same way, each element of the basis is labelled by pairs of permutation lattices.
\par
As in the case of subduction problem for $\mathfrak{S}_f$, we write the non-standard base vectors $|f_1, f_2; \lambda_1, \lambda_2; w_1,w_2\rangle$ of $[f_1, \lambda_1]\otimes[f_2, \lambda_2]$ in terms of the standard base vectors $|f; \lambda; w\rangle$ of $[f,\lambda]$ ($f_1 +f_2 = f$):
\begin{equation}
|f_1, f_2; \lambda_1, \lambda_2; w_1,w_2\rangle_{\eta} = \sum_{w\in \Xi_f^{\lambda}} \ |f; \lambda; w\rangle\langle f; \lambda; w|f_1, f_2; \lambda_1, \lambda_2; w_1, w_2 \rangle_{\eta}
\end{equation}
where $\Xi_f^{\lambda}$ represents the set of all permutation lattices of order $f$ and diagram $\lambda$. Thus $\langle f; \lambda; w|f_1, f_2; \lambda_1, \lambda_2; w_1, w_2 \rangle_{\eta}$ are the SDCs of $[f,\lambda]\downarrow[f_1,\lambda_1]\otimes[f_2,\lambda_2]$ with given multiplicity label $\eta$.
\par 
Again, the SDCs satisfy the following unitary conditions:
\begin{equation}
\sum_w \ \langle f; \lambda ; w | f_1, f_2;\lambda_1, \lambda_2 ; w_1, w_2 \rangle_{\eta} \ \langle f; \lambda ; w | f_1, f_2;\lambda_1, \lambda'_2 ; w_1, w'_2 \rangle_{\eta'} = \delta_{\lambda_2 \lambda'_2} \delta_{w_2 w'_2} \delta_{\eta \eta'}
\label{orton1br}
\end{equation}
\begin{equation}
\sum_{\lambda_2 w_2 \eta} \ \langle f;\lambda ; w |f_1, f_2; \lambda_1, \lambda_2 ; w_1, w_2 \rangle_{\eta} \ \langle f; \lambda ; w' |f_1,f_2; \lambda_1, \lambda_2 ; w_1, w_2 \rangle_{\eta} = \delta_{w w'}.
\label{orton2br}
\end{equation}
\subsection{Subduction system}
Following the guidelines given for the subduction problem in symmetric groups, we now construct a matrix in such a way that the SDCs are the components of the kernel basis vectors. The dimension of such a kernel space is equal to the multiplicity for the subduction issue we are considering.
\par 
The action of $g_i$ and $e_i$ on the non-standard base vectors is given by
\begin{equation}
g_i |f_1, f_2; \lambda_1, \lambda_2 ; w_1, w_2 \rangle=
\left\{
\begin{array}{cc}
 (g_i |f_1;\lambda_1; w_1 \rangle) \otimes |f_2;\lambda_2; w_2 \rangle & \text{if $1 \le i \le f_{1}-1$ } \\
|f_1;\lambda_1; w_1 \rangle \otimes (g_i|f_2;\lambda_2; m_2 \rangle) & \text{if $f_1+1 \le i \le f-1$}
\end{array}
\right.
\label{actsplitbrg} 
\end{equation}
and 
\begin{equation}
e_i |f_1, f_2; \lambda_1, \lambda_2 ; w_1, w_2 \rangle=
\left\{
\begin{array}{cc}
 (e_i |f_1;\lambda_1; w_1 \rangle) \otimes |f_2;\lambda_2; w_2 \rangle & \text{if $1 \le i \le f_{1}-1$ } \\
|f_1;\lambda_1; w_1 \rangle \otimes (e_i|f_2;\lambda_2; m_2 \rangle) & \text{if $f_1+1 \le i \le f-1$}
\end{array}.
\right.
\label{actsplitbre} 
\end{equation}
From (\ref{actsplitbrg}) and (\ref{actsplitbre}), for $1 \leq l \leq f_1 - 1  $, we get 
\begin{equation}
\langle f; \lambda ; w | g_l | f_1, f_2; \lambda_1, \lambda_2; w_1, w_2 \rangle = \langle f; \lambda ; w | (g_l | f_1;\lambda_1; w_1 \rangle) \otimes | f_2; \lambda_2; w_2 \rangle
\label{r1gbr}
\end{equation}
and
\begin{equation}
\langle f; \lambda ; w | e_l | f_1, f_2; \lambda_1, \lambda_2; w_1, w_2 \rangle = \langle f; \lambda ; w | (e_l | f_1;\lambda_1; w_1 \rangle) \otimes | f_2; \lambda_2; w_2 \rangle.
\label{r1ebr}
\end{equation}
Writing $| f_1, f_2; \lambda_1, \lambda_2; w_1, w_2 \rangle$ and $g_l | f_1;\lambda_1; w_1 \rangle$ in the standard basis of $[f,\lambda]$ and $[f_1, \lambda_1]$ respectively, (\ref{r1gbr}) and (\ref{r1ebr}) become
\begin{multline}
\sum_{u\in \Theta_i(w)} \ \langle f; \lambda; w | g_l | f; \lambda; u \rangle \langle f; \lambda ; u |f_1, f_2 ; \lambda_1, \lambda_2 ; w_1, w_2 \rangle  =  \\ \sum_{v\in \Theta_i(w_1)} \ \langle f_1; \lambda_1; v | g_l |f_1; \lambda_1; w_1  \rangle \langle f; \lambda ; w | f_1,f_2; \lambda_1, \lambda_2 ; v , w_2 \rangle
\label{lem1gbr}
\end{multline}
\begin{multline}
\sum_{u \in \bar{\Theta}_i(w)} \ \langle f; \lambda; w | e_l | f; \lambda; u \rangle \langle f; \lambda ; u |f_1, f_2 ; \lambda_1, \lambda_2 ; w_1, w_2 \rangle  =  \\ \sum_{v \in \bar{\Theta}_i(w_1)} \ \langle f_1; \lambda_1; v | e_l |f_1; \lambda_1; w_1  \rangle \langle f; \lambda ; w | f_1,f_2; \lambda_1, \lambda_2 ; v , w_2 \rangle
\label{lem1ebr}
\end{multline}
where $\Theta_i(w)$ and $\bar{\Theta}_i(w)$ denote the sets of all permutation lattices which are $i$-coupled and $\bar{i}$-coupled with $w$ respectively. 
\par 
In an analogous way, for $f_1+1 \leq l \leq f-1$, we get
\begin{multline}
\sum_{u\in \Theta_i(w)} \ \langle f; \lambda; w | g_l | f; \lambda; u \rangle \langle f; \lambda ; u |f_1, f_2 ; \lambda_1, \lambda_2 ; w_1, w_2 \rangle  =  \\ \sum_{v\in \Theta_i(w_2)} \ \langle f_2; \lambda_2; v | g_l |f_2; \lambda_2; w_2  \rangle \langle f; \lambda ; w | f_1,f_2; \lambda_1, \lambda_2 ; w_1 , v \rangle
\label{lem2gbr}
\end{multline}
\begin{multline}
\sum_{u \in \bar{\Theta}_i(w)} \ \langle f; \lambda; w | e_l | f; \lambda; u \rangle \langle f; \lambda ; u |f_1, f_2 ; \lambda_1, \lambda_2 ; w_1, w_2 \rangle  =  \\ \sum_{v \in \bar{\Theta}_i(w_2)} \ \langle f_2; \lambda_2; v | e_l |f_2; \lambda_2; w_2  \rangle \langle f_1,f_2; \lambda ; w | f_1; \lambda_1, \lambda_2 ; w_1 , v \rangle.
\label{lem2ebr}
\end{multline}
Then, once we know the explicit action of the generators of $\mathfrak{B}_{f_1}(x) \times \mathfrak{B}_{f_2}(x)$ on the standard basis, (\ref{lem1gbr}), (\ref{lem1ebr}), (\ref{lem2gbr}) and (\ref{lem2ebr}) (written for all $l\in\{1, \ldots, f_1-1, f_1+1, \ldots, f-1 \}$ and all permutation lattices $w$, $w_1$, $w_2$ of order $f$ and diagrams $\lambda$, $\lambda_1$ and $\lambda_2$ respectively) define a linear equation system of the form:
\begin{equation}
\Omega(\lambda; f_1, f_2; \lambda_1, \lambda_2) \ \chi = 0
\label{subdmbr}
\end{equation}
where $\Omega(\lambda; f_1, f_2; \lambda_1, \lambda_2)$ is the \emph{subduction matrix} and $\chi$ is a vector with components given by the SDCs of $[f,\lambda]\downarrow[f_1, \lambda_1]\otimes[f_2, \lambda_2]$. (\ref{subdmbr}) is a linear equation system with $\dim(\mathfrak{B}_f(x); [f,\lambda]) \cdot \dim(\mathfrak{B}_{f_1}(x); [f_1,\lambda_1]) \cdot \dim(\mathfrak{B}_{f_2}(x); [f_2,\lambda_2])$ unknowns (the SDCs) and $2(f-2)\cdot\dim(\mathfrak{B}_f(x); [f,\lambda])\cdot \dim(\mathfrak{B}_{f_1}(x); [f_1,\lambda_1])\cdot \dim(\mathfrak{B}_{f_2}(x); [f_2,\lambda_2]) $ equations. 
\subsection{Explicit form of the subduction system}
It will be useful to give the following definitions of $i$-coupling and $\bar{i}$-coupling on pairs of permutation lattices:
\begin{definition}
Given two pairs $w_{12}=(w_1,w_2)$ and $w'_{12}=(w'_1,w'_2)$, each one composed of two permutation lattices of order $f_1$ and $f_2$ respectively, we say that $w_{12}$ is \emph{$i$-coupled} to $w'_{12}$ (or that $w_{12}$ and $w'_{12}$ are \emph{$i$-coupled}) when 
\begin{equation}
\left\{
\begin{array}{cc}
\begin{array}{c}
w_1 \stackrel{i}\leftrightarrow w'_1 \\ 
w_2=w'_2
\end{array} & \textmd{if $1 \leq i \leq f_1-1$} \\
 & \\
\begin{array}{c}
w_1 = w'_1 \\
w_2 \stackrel{i-f_1}\leftrightarrow w'_2
\end{array}
 & \textmd{if $f_1 +1 \leq i \leq f_1 + f_2 -1$}
\end{array}
\right.
\nonumber
\end{equation}
and we denote such a relation by $w_{12}\stackrel{i}\leftrightarrow w'_{12}$.
\end{definition}
\begin{definition}
Given two pairs $w_{12}=(w_1,w_2)$ and $w'_{12}=(w'_1,w'_2)$, each one composed of two permutation lattices of order $f_1$ and $f_2$ respectively, we say that $w_{12}$ is \emph{$\bar{i}$-coupled} to $w'_{12}$ (or that $w_{12}$ and $w'_{12}$ are \emph{$\bar{i}$-coupled}) when 
\begin{equation}
\left\{
\begin{array}{cc}
\begin{array}{c}
w_1 \stackrel{\bar{i}}\leftrightarrow w'_1 \\ 
w_2=w'_2
\end{array} & \textmd{if $1 \leq i \leq f_1-1$} \\
 & \\
\begin{array}{c}
w_1 = w'_1 \\
w_2 \stackrel{\overline{i-f_1}}\leftrightarrow w'_2
\end{array}
 & \textmd{if $f_1 +1 \leq i \leq f_1 + f_2 -1$}
\end{array}
\right.
\nonumber
\end{equation}
and we denote such a relation by $w_{12}\stackrel{\bar{i}}\leftrightarrow w'_{12}$.
\end{definition}
Of course, an $(i\setminus\bar{i})$-coupling relation on pairs of permutation lattices can be also defined by
\begin{equation}
w_{12} \stackrel{i\setminus\bar{i}}\leftrightarrow w'_{12} \Longleftrightarrow 
\left\{
\begin{array}{cc}
\begin{array}{c}
w_1 \stackrel{i\setminus\bar{i}}\leftrightarrow w'_1 \\ 
w_2=w'_2
\end{array} & \textmd{if $1 \leq i \leq f_1-1$} \\
 & \\
\begin{array}{c}
w_1 = w'_1 \\
w_2 \stackrel{i\setminus\overline{i-f_1}}\longleftrightarrow w'_2
\end{array}
 & \textmd{if $f_1 +1 \leq i \leq f_1 + f_2 -1$}
\end{array}
\right.
\nonumber
\end{equation}
\par
Denoted by $\Theta_i(w_{12})$ the set of all pairs of permutation lattices which are $i$-coupled to the pair $w_{12}=(w_1,w_2)$ and by $\bar{\Theta}_{i}(w_{12})$ the set of all pairs of permutation lattices which are $\bar{i}$-coupled to the pair $w_{12}=(w_1,w_2)$, equations (\ref{lem1gbr}), (\ref{lem1ebr}), (\ref{lem2gbr}) and  (\ref{lem2ebr}) can be written as
\begin{multline}
(\langle f_1, f_2; \lambda_1, \lambda_2; w_1, w_2 | g_i  |f_1, f_2;\lambda_1, \lambda_2; w_1, w_2 \rangle - \langle f; \lambda; w | g_i  |f;\lambda; w \rangle) \ - \\
\sum_{
u\in \Theta'_i(w)} 
\  \langle f; \lambda; w | g_i  |f;\lambda; u \rangle  \langle f; \lambda ; u |f_1, f_2 ; \lambda_1, \lambda_2 ; w_1, w_2 \rangle \ + 
\sum_{
(u_1, u_2)\in \Theta'_i(w_{12})} \\ \langle f_1, f_2; \lambda_1, \lambda_2; w_1, w_2 | g_i  |f_1, f_2;\lambda_1, \lambda_2; u_1, u_2 \rangle \langle f; \lambda ; w |f_1, f_2 ; \lambda_1, \lambda_2 ; u_1, u_2 \rangle = 0
\label{eqg}
\end{multline}
and
\begin{multline}
(\langle f_1, f_2; \lambda_1, \lambda_2; w_1, w_2 | e_i  |f_1, f_2;\lambda_1, \lambda_2; w_1, w_2 \rangle - \langle f; \lambda; w | e_i  |f;\lambda; w \rangle) \ - \\
\sum_{
u\in \bar{\Theta}'_i(w)} 
\  \langle f; \lambda; w | e_i  |f;\lambda; u \rangle  \langle f; \lambda ; u |f_1, f_2 ; \lambda_1, \lambda_2 ; w_1, w_2 \rangle \ +  
\sum_{
(u_1, u_2)\in \bar{\Theta}'_i(w_{12})} \\ \langle f_1, f_2; \lambda_1, \lambda_2; w_1, w_2 | e_i  |f_1, f_2;\lambda_1, \lambda_2; u_1, u_2 \rangle \langle f; \lambda ; w |f_1, f_2 ; \lambda_1, \lambda_2 ; u_1, u_2 \rangle  =  0,
\label{eqe}
\end{multline}
where  $\Theta'_i(w)$ and $\Theta'_{\bar{i}}(w)$ represent the sets $\Theta_i(w)\setminus\{w\}$ and $\Theta_{\bar{i}}(w)\setminus\{w\}$, respectively (and analogously for  $\Theta'_i(w_{12})$ and $\Theta'_{\bar{i}}(w_{12})$).
\par 
By remembering the statement of theorem~\ref{action}, we can distinguish four possible cases for the structure of the equations (\ref{eqg}) and (\ref{eqe}).
\begin{enumerate}
\item \emph{Crossing}: $w\stackrel{i\setminus\bar{i}}\leftrightarrow w$ and $w_{12}\stackrel{i\setminus\bar{i}}\leftrightarrow w_{12}$. \\
The subduction equations become of the form given in~\cite{chilla}:
\begin{equation}
\alpha^{(i\setminus\bar{i})}_{w,w_{12}} \langle f; \lambda;  w |f_1, f_2; \lambda_1, \lambda_2 ; w_1, w_2 \rangle - \beta^{(i\setminus\bar{i})}_w  \langle f; \lambda ;  g_i(w) | f_1, f_2; \lambda_1, \lambda_2 ; w_1, w_2 \rangle + 
\nonumber
\end{equation}
\begin{equation}
+ \beta^{(i\setminus\bar{i})}_{w_{12}} \langle f; \lambda ;  w| f_1, f_2;\lambda_1 , \lambda_2 ; g_i(w_1),w_2 \rangle = 0 \ \ \ \ \ \ \  \text{if $i \in \{ 1, \ldots, n_1 - 1 \}$ },
\end{equation}
\begin{equation}
\alpha^{(i\setminus\bar{i})}_{w,w_{12}} \langle f; \lambda;  w | f_1, f_2;\lambda_1, \lambda_2 ; w_1, w_2 \rangle - \beta^{(i\setminus\bar{i})}_w  \langle f; \lambda ;  g_i(w) |f_1, f_2; \lambda_1, \lambda_2 ; w_1, w_2 \rangle + 
\nonumber
\end{equation}
\begin{equation}
+ \beta^{(i\setminus\bar{i})}_{w_{12}} \langle f; \lambda ;  w| f_1, f_2; \lambda_1 , \lambda_2 ; w_1,g_i(w_2) \rangle = 0 \ \ \ \ \ \ \  \text{if $i \in \{ n_1 + 1, \ldots, n - 1 \}$ }
\label{sueqrid1}
\end{equation}
where 
\begin{equation}
 \alpha^{(i\setminus\bar{i})}_{w,w_{12}} = \frac{1}{d_i(w_{12})} - \frac{1}{d_i(w)} 
\end{equation}
\begin{equation}
\beta^{(i\setminus\bar{i})}_w = \sqrt{1-\frac{1}{d_i^2(w)}}
\end{equation}
\begin{equation}
\beta^{(i\setminus\bar{i})}_{w_{12}} = \sqrt{1-\frac{1}{d_i^2(w_{12})}}.
\end{equation}
Notice that, by definition, 
\begin{equation}
d_i(w_{12}) =
\left\{
\begin{array}{cc}
d_i(w_1) & \text{if $1\leq i \leq f_1-1$} \\  
d_{i-f_1}(w_2) & \text{if $f_1 + 1 \leq i \leq f-1$}
\end{array}
\right. ,
\end{equation}
where the axial distance $d_i$ is the same of (\ref{dass}) and, given a permutation lattice $w=(w_1, \ldots, w_i, w_{i+1}, \ldots, w_f)$, the $g_i$ action is naturally defined in the following way: consider the word $\tilde{w}=(w_1, \ldots,w_{i-1,} w_{i+1,} w_{i}, w_{i+2}, \ldots, w_f)$ obtained by $w$ interchanging the elements $w_i$ and $w_{i+1}$. If $\tilde{w}$ is another permutation lattice then we put $g_i(w)=\tilde{w}$, otherwise we set $g_i(w)=w$. In an analogous way, it is defined a $g_i$ action on pairs of permutation lattices of order $f_1$ and $f_2$, respectively:
\begin{equation}
g_i(w_1,w_2)=
\left\{
\begin{array}{cc}
(g_i(w_1), w_2) & \textmd{if $1 \leq i \leq f_1-1$} \\
(w_1, g_i(w_2)) & \textmd{if $f_1+1 \leq i \leq f_1 + f_2 -1$}
\end{array}.
\right. 
\end{equation}
\item \emph{Horizontal bridge}: $w\stackrel{\bar{i}}\leftrightarrow w$ and $w_{12}\stackrel{i\setminus\bar{i}}\leftrightarrow  w_{12}$.  \\
In this case, we get the equations:
\begin{multline}
\alpha^{(i\setminus\bar{i})}_{w,w_{12}} \langle f; \lambda;  w |f_1, f_2; \lambda_1, \lambda_2 ; w_{12} \rangle + \beta^{(i\setminus\bar{i})}_{w_{12}} \langle \lambda;  w |f_1, f_2; \lambda_1, \lambda_2 ; g_i(w_{12})\rangle = \\
- \frac{\sqrt{P_{Y(w^{(i)})}(x)}}{P_{Y(w^{(i-1)})}(x)} \sum_{u\in\bar{\Theta}_{i}(w)} \ \frac{\sqrt{P_{Y(u^{(i)})}(x)}}{\diamondsuit_i(w,u)} \langle f; \lambda;  u |f_1, f_2; \lambda_1, \lambda_2 ; w_{12} \rangle 
\label{ghb}
\end{multline}
and
\begin{equation}
\sum_{u\in\bar{\Theta}_{i}(w)}\sqrt{P_{Y(u^{(i)})}(x)} \langle f; \lambda;  u |f_1, f_2; \lambda_1, \lambda_2 ; w_{12} \rangle = 0 ,
\label{kerhb}
\end{equation}
where we have used the usual notation $w_{12}=(w_1, w_2)$ in the mathematical symbol of SDC.
\item \emph{Vertical bridge}: $w\stackrel{i\setminus\bar{i}}\leftrightarrow w$ and $w_{12}\stackrel{\bar{i}}\leftrightarrow w_{12}$. \\
In an analogous way as the previous case, we have:
\begin{multline}
\alpha^{(i\setminus\bar{i})}_{w,w_{12}} \langle f; \lambda;  w |f_1, f_2; \lambda_1, \lambda_2 ; w_{12} \rangle - \beta^{(i\setminus\bar{i})}_{w} \langle \lambda;  g_i(w) |f_1, f_2; \lambda_1, \lambda_2 ; w_{12} \rangle = \\
\frac{\sqrt{P_{Y(w_{12}^{(i)})}(x)}}{P_{Y(w_{12}^{(i-1)})}(x)} \sum_{u_{12}\in\bar{\Theta}_{i}(w_{12})} \ \frac{\sqrt{P_{Y(u_{12}^{(i)})}(x)}}{\diamondsuit_i(w_{12},u_{12})} \langle f; \lambda;  w |f_1, f_2; \lambda_1, \lambda_2 ; u_{12} \rangle 
\label{gvb}
\end{multline}
and
\begin{equation}
\sum_{u_{12}\in\bar{\Theta}_{i}(w_{12})} \ \sqrt{P_{Y(u_{12}^{(i)})}(x)} \langle f; \lambda;  u |f_1, f_2; \lambda_1, \lambda_2 ; w_{12} \rangle = 0 .
\label{kervb}
\end{equation}
Here, as in the definition of axial distance for pairs of permutation lattices, we have set:
\begin{equation}
\diamondsuit_i(w_{12}, u_{12}) = 
\left\{
\begin{array}{cc}
\diamondsuit_i(w_1, u_1) & \textmd{if $1 \leq i \leq f_1-1 $} \\
\diamondsuit_{i-f_1}(w_2, u_2) & \textmd{if $f_1+1 \leq i \leq f-1$}
\end{array}
\right. 
\end{equation}
and
\begin{equation}
w_{12}^{(i)} = 
\left\{
\begin{array}{cc}
w_1^{(i)} & \textmd{if $1 \leq i \leq f_1-1 $} \\
w_2^{(i-f_1)} & \textmd{if $f_1+1 \leq i \leq f-1$}
\end{array}
\right. .
\end{equation}
\item \emph{Singlet}: $w\stackrel{\bar{i}}\leftrightarrow w$ and $w_{12}\stackrel{\bar{i}}\leftrightarrow w_{12}$. \\
In this last case, the subduction equations take the form of:
\begin{multline}
\alpha^{(i\setminus\bar{i})}_{w,w_{12}} \langle f; \lambda;  w |f_1, f_2; \lambda_1, \lambda_2 ; w_{12} \rangle = \\
- \frac{\sqrt{P_{Y(w^{(i)})}(x)}}{P_{Y(w^{(i-1)})}(x)} \sum_{u\in\bar{\Theta}_{i}(w)} \ \frac{\sqrt{P_{Y(u^{(i)})}(x)}}{\diamondsuit_i(w,u)} \langle f; \lambda;  u |f_1, f_2; \lambda_1, \lambda_2 ; w_{12} \rangle + \\
\frac{\sqrt{P_{Y(w_{12}^{(i)})}(x)}}{P_{Y(w_{12}^{(i-1)})}(x)} \sum_{u_{12}\in\bar{\Theta}_{i}(w_{12})} \ \frac{\sqrt{P_{Y(u_{12}^{(i)})}(x)}}{\diamondsuit_i(w_{12},u_{12})} \langle f; \lambda;  w |f_1, f_2; \lambda_1, \lambda_2 ; u_{12} \rangle 
\end{multline}
and
\begin{multline}
\frac{\sqrt{P_{Y(w^{(i)})}(x)}}{P_{Y(w^{(i-1)})}(x)} \sum_{u\in\bar{\Theta}_{i}(w)} \ \sqrt{P_{Y(u^{(i)})}(x)} \langle f; \lambda;  u |f_1, f_2; \lambda_1, \lambda_2 ; w_{12} \rangle = \\
\frac{\sqrt{P_{Y(w_{12}^{(i)})}(x)}}{P_{Y(w_{12}^{(i-1)})}(x)} \sum_{u_{12}\in\bar{\Theta}_{i}(w_{12})} \ \sqrt{P_{Y(u_{12}^{(i)})}(x)}\langle f; \lambda;  w |f_1, f_2; \lambda_1, \lambda_2 ; u_{12} \rangle.
\label{intere}
\end{multline}
\end{enumerate}


\section{Subduction graph}
Let us now consider the three shapes $(f; \lambda; f_1, f_2; \lambda_1, \lambda_2)$ with $f_1 + f_2 = f$. We call \emph{node} each ordered sequence of three permutation lattices $(w; w_1,w_2)$ such as $w \in \Xi_f^{\lambda}$, $w \in \Xi_{f_1}^{\lambda_1}$ and $w \in \Xi_{f_2}^{\lambda_2}$. We denote it by $\langle w; w_1, w_2 \rangle$ or $\langle w; w_{12} \rangle$. The set of all nodes of $(f; \lambda; f_1, f_2; \lambda_1, \lambda_2)$ is called \emph{subduction grid} (or simply \emph{grid}) and it is as usual denoted by $G$. Building on the case of permutation lattices, the following definition extends the $i$-coupling relation to the nodes of the grid.
\begin{definition}
Fixed the grid $(f;\lambda;f_1, f_2; \lambda_1, \lambda_2)$, and given two nodes $n=(w;w_{12})$ and $n'=(w';w'_{12})$, we say that $n$ is \emph{$i$-coupled} to $n'$ (or that $n$ and $n'$ are \emph{$i$-coupled}) if $w\stackrel{i}\leftrightarrow w'$ and $w_{12}\stackrel{i}\leftrightarrow w'_{12}$. Then we write $n\stackrel{i}\leftrightarrow n'$. 
\end{definition}
Once $i$ is fixed, it easy to see that the $i$-coupling is an equivalence relation on the grid. We name $i$-layer the partition of $G$ which is associated to such a relation and we denote it by $G^{(i)}$. If $n$ and $n'$ are two distinct nodes of the grid such that $n\stackrel{i}\leftrightarrow n'$, then they are connected by an edge with a label for $i$.
\par 
Following the structure of explicit form for the subduction equations given in the previous section, we note that there are only four possible kinds of \emph{$i$-layer configurations} beetween nodes in $G$:
\begin{enumerate}
\item \emph{crossing} $i$-layer: $G^{(i\setminus\bar{i})}=\{ \langle w; w_{12} \rangle\in G \ | \ w\stackrel{i\setminus\bar{i}}\leftrightarrow w \textmd{ and } w_{12}\stackrel{i\setminus\bar{i}}\leftrightarrow w_{12} \}$;
\item \emph{horizontal bridge} $i$-layer: $G^{(i-\bar{i})}=\{ \langle w; w_{12} \rangle\in G \ | \ w\stackrel{\bar{i}}\leftrightarrow w \textmd{ and } w_{12}\stackrel{i\setminus\bar{i}}\leftrightarrow w_{12} \}$;
\item \emph{vertical bridge} $i$-layer: $G^{(\bar{i}-i)}=\{ \langle w; w_{12} \rangle\in G \ | \ w\stackrel{i\setminus\bar{i}}\leftrightarrow w \textmd{ and } w_{12}\stackrel{\bar{i}}\leftrightarrow w_{12} \}$;
\item \emph{singlet} $i$-layer: $G^{(\bar{i})}=\{ \langle w; w_{12} \rangle\in G \ | \ w\stackrel{\bar{i}}\leftrightarrow w \textmd{ and } w_{12}\stackrel{\bar{i}}\leftrightarrow w_{12} \}$;
\end{enumerate}
Clearly, $G^{(i\setminus\bar{i})}$, $G^{(i-\bar{i})}$, $G^{(\bar{i}-i)}$ and $G^{(\bar{i})}$ are disjoint sets and we have $G^{(i)}= G^{(i\setminus\bar{i})} \cup G^{(i-\bar{i})} \cup G^{(\bar{i}-i)} \cup G^{(\bar{i})}$. The crossing $i$-layer corresponds to the $i$-layer defined in~\cite{chilla} for the subduction problem in symmetric groups. So, crossing, bridge and singlet configurations for the $i$-coupling relation are also defined in an analogous way for such a set.
\begin{definition} 
We call \emph{subduction graph} the \emph{overlap} of all $i$-layers obtained by identification of the corresponding nodes.
\end{definition}
The definition just given is a good definition of subduction graph, because there is at most \emph{one} edge connecting two distinct nodes. This is ensured by the osservation that if $n$ and $n'$ are two distinct nodes which are $i$-coupled and $j$-coupled then we necessarily have $i=j$.  
\par 
We remark that if the grid defined by $(f;\lambda;f_1, f_2; \lambda_1, \lambda_2)$ is such that $f$ is equal to the number of boxes of $\lambda$, $f_1$ is equal to the number of boxes of $\lambda_1$ and $f_3$ to the number of boxes $\lambda_3$, then the definition of subduction graph just given becomes that one given for the subduction problem in symmetric groups. 


\section{Structure of the subduction space}
The solution of (\ref{subdmbr}) can be seen as an intersection of $f-2$ subspaces $\chi^{(i)}$ such that each one satisfies 
\begin{equation}
\Omega^{(i)}(f_1, f_2; \lambda; \lambda_1, \lambda_2) \ \chi^{(i)}= 0. 
\label{nullspbr}
\end{equation}
Here, $\Omega^{(i)}(f_1, f_2; \lambda; \lambda_1, \lambda_2)$ is defined by the equations (\ref{lem1gbr}), (\ref{lem1ebr}), (\ref{lem2gbr}) and (\ref{lem2ebr}) written for a fixed $i\in \{ 1, \ldots, f_1-1,f_1+1, \ldots f-1\}$.
The definitions of grid, $i$-layer and the explicit form for the subduction equations, given in the previous sections, provide a suitable way to describe the solution space of (\ref{nullspbr}) by the one-to-one corrispondence between the nodes of $(f_1, f_2; \lambda; \lambda_1, \lambda_2)$ and the SDCs for the subduction $[f_1+f_2,\lambda]\downarrow [f_1,\lambda_1]\otimes [f_2, \lambda_2]$. To find the structure of the subduction space $\chi^{(i)}$ which is associated to the $i$-layer we only need to describe the structure of the spaces which are associated to $G^{(i\setminus\bar{i})}$, $G^{(i-\bar{i})}$, $G^{(\bar{i}-1)}$ and $G^{(\bar{i})}$ that we call \emph{crossing space}, \emph{horizontal bridge space}, \emph{vertical bridge space} and \emph{singlet space}, respectively.
\subsection{Crossing space}
The solution for the crossing equations was already described in~\cite{chilla, chilla1} by the subduction graph method. In fact, we observe that the structure of the two subduction systems are quite similar. On the other hand, the irreps of $\mathfrak{S}_f$ also are irreps of $\mathfrak{B}_f(x)$. For the Brauer algebras case, we only need to pay attention to use the new definition of axial distance given in (\ref{dass}) because such a definition leads to expressions for the coefficients $\alpha^{(i\setminus\bar{i})}_{w;w_{12}}$, $\beta^{(i\setminus\bar{i})}_w$ and $\beta^{(i\setminus\bar{i})}_{w_{12}}$ which are algebraic functions of $\mathbb{C}(x)$ instead of simple real numbers (see theorem~\ref{action}). However, relations and conditions of the subduction graph method for symmetric groups still remain valid for the Brauer algebras subduction issue.  
\subsection{Bridge spaces}
Let us now consider the case of the horizontal bridge space. From equation (\ref{kerhb}), for each node $\langle w;w_{12} \rangle \in G^{(i-\bar{i})}$, we find that subduction coefficients of the horizontal bridge type, $\langle f; \lambda; u | f_1, f_2; \lambda_1, \lambda_2; w_{12} \rangle$, are the components of vectors of a vectorial space that is the kernel of the operator $e_i$ acting on the invariant irreducible subspace defined by all the permutation lattices $u$ which are $\bar{i}$-coupled to $w$. From the relation $e_i^2 = x e_i$, we note that the eigenvalues of $e_i$ are $0$ and $x$. Therefore $\chi^{(i-\bar{i})}$ in general is not the trivial space. So, finding such SDCs is equivalent to determind the kernel space of $e_i$ in the explicit form given in theorem~\ref{action}.
\par  
Once we now the SDCs of the form $\langle f; \lambda; u | f_1, f_2; \lambda_1, \lambda_2; w_{12} \rangle$, we can determine the coefficients $\langle f; \lambda; u | f_1, f_2; \lambda_1, \lambda_2; g_i(w_{12}) \rangle$ by using (\ref{ghb}):
\begin{multline}
\langle f; \lambda; u | f_1, f_2; \lambda_1, \lambda_2; g_i(w_{12}) \rangle =  - \frac{\alpha^{(i\setminus\bar{i})}_{w,w_{12}}}{\beta_{w_{12}}^{(i\setminus\bar{i})}} \langle f; \lambda;  w |f_1, f_2; \lambda_1, \lambda_2 ; w_{12} \rangle - \\ \frac{1}{\beta_{w}^{(i\setminus\bar{i})}}\frac{\sqrt{P_{Y(w^{(i)})}(x)}}{P_{Y(w^{(i-1)})}(x)} \sum_{u\in\bar{\Theta}_{i}(w)} \ \frac{\sqrt{P_{Y(u^{(i)})}(x)}}{\diamondsuit_i(w,u)} \langle f; \lambda;  u |f_1, f_2; \lambda_1, \lambda_2 ; w_{12} \rangle 
\end{multline}
(note that if $g_i(w_{12})\neq w_{12}$ then $\beta_{w_{12}}^{(i\setminus\bar{i})} \neq 0$).
\par 
In an analogous way for the vertical bridge space, from equation (\ref{kervb}) we find that subduction coefficients $\langle f; \lambda; w | f_1, f_2; \lambda_1, \lambda_2; u_{12} \rangle$, are the components of vectors of a vectorial space that is the kernel of the operator $e_i$ acting on the invariant irreducible subspace defined by all pairs of the permutation lattices $u_{12}$ which are $\bar{i}$-coupled to $w_{12}$.  
\par 
Again, once we now the SDCs of the form $\langle f; \lambda; w | f_1, f_2; \lambda_1, \lambda_2; u_{12} \rangle$, we can determine the coefficients $\langle f; \lambda; u | f_1, f_2; \lambda_1, \lambda_2; g_i(w_{12}) \rangle$ by using (\ref{gvb}):
\begin{multline}
\langle f; \lambda; w | f_1, f_2; \lambda_1, \lambda_2; g_i(u_{12}) \rangle =  \frac{\alpha^{(i\setminus\bar{i})}_{w,w_{12}}}{\beta_{w}^{(i\setminus\bar{i})}} \langle f; \lambda;  w |f_1, f_2; \lambda_1, \lambda_2 ; w_{12} \rangle - \\ \frac{1}{\beta_{w}^{(i\setminus\bar{i})}}\frac{\sqrt{P_{Y(w_{12}^{(i)})}(x)}}{P_{Y(w_{12}^{(i-1)})}(x)} \sum_{u_{12}\in\bar{\Theta}_{i}(w_{12})} \ \frac{\sqrt{P_{Y(u_{12}^{(i)})}(x)}}{\diamondsuit_i(w_{12},u_{12})} \langle f; \lambda;  w |f_1, f_2; \lambda_1, \lambda_2 ; u_{12} \rangle 
\end{multline}
(if $g_i(w)\neq w$ then $\beta_{w}^{(i\setminus\bar{i})} \neq 0$.)
\subsection{Singlet space}
To understand the structure of the solution for singlet equations, it is useful to introduce the \emph{intertwining} operators:
\begin{equation}
{\Omega}_{w, w_{12}}^{(i)} = I_w \otimes {\rho}_{w_{12}}^{(i)} - {\rho}_w^{(i)} \otimes I_{w_{12}}.
\end{equation}
and 
\begin{equation}
\bar{\Omega}_{w, w_{12}}^{(i)} = I_w \otimes \bar{\rho}_{w_{12}}^{(i)} - \bar{\rho}_w^{(i)} \otimes I_{w_{12}}
\end{equation}
Here, ${\rho}_w^{(i)}$ (resp. ${\rho}_{w_{12}}^{(i)}$) represents the action of the generators $g_i$ on the invariant irreducible module defined by all permutation lattices (resp. pairs of permutation lattices) which are $\bar{i}$-coupled to $w$ (risp. $w_{12}$). In an analogous way, $\bar{\rho}_w^{(i)}$ (resp. $\bar{\rho}_{w_{12}}^{(i)}$) represents the action of the generators $e_i$ on the invariant irreducible modules defined by all permutation lattices (resp. pairs of permutation lattices) which are $\bar{i}$-coupled to $w$ (resp. $w_{12}$). Furthermore, $I_w$ and $I_{w_{12}}$ represent the identity operators on the previous invariant irreducible modules, respectively. Solving the singlet equations is equivalent to find the kernel space of ${\Omega}_{w, w_{12}}^{(i)}$ and the kernel space of $\bar{\Omega}_{w, w_{12}}^{(i)}$.
\par   
The operator $\rho_w^{(i)}$ (resp. $\rho_{w_{12}}^{(i)}$) has eigenvalues $1$ and $-1$, as we can see by the relation $g_i^2=1$. Denoted by $g_{w,1}^{(i)}$ (resp. $g_{{w_{12}},1}^{(i)}$) the eigenvector relative to the eigenvalue $1$ and by $g_{w,-1}^{(i)}$ (resp. $g_{{w_{12}},-1}^{(i)}$ ) that one relative to the eigenvalue $-1$, the eigenvectors of the intertwining operator $\Omega_{w, w_{12}}^{(i)}$ are: 
\begin{enumerate}
\item $g_{w,1}^{(i)}\otimes g_{{w_{12}},1}^{(i)} $ with eigenvalue $0$;
\item ${g}_{w,-1}^{(i)}\otimes g_{{w_{12},1}}^{(i)}$ with eigenvalue $2$;
\item $g_{w,1}^{(i)}\otimes{g}_{{w_{12}},-1}^{(i)}$ with eigenvalue $-2$;
\item ${g}_{w,-1}^{(i)}\otimes {g}_{{w_{12}},-1}^{(i)}$ with eigenvalue $0$.
\end{enumerate}
Therefore, the kernel space is given by $\spanvettoriale(g_{w,1}^{(i)}\otimes g_{{w_{12}},1}^{(i)},{g}_{w,-1}^{(i)}\otimes {g}_{{w_{12}},-1}^{(i)})$.
\par 
The operator $\bar{\rho}_w^{(i)}$ (resp. $\bar{\rho}_{w_{12}}^{(i)}$) has eigenvalues $x$ and $0$ (remember that ${e_i}^2=x e_i$). Denoted by $e_{w,x}^{(i)}$ (resp. $e_{{w_{12}},x}^{(i)}$) an eigenvector relative to the eigenvalue $x$ and by $e_{w,0}^{(i)}$ (resp. $e_{{w_{12}},0}^{(i)}$) one relative to the eigenvalue $-1$, the eigenvectors of the intertwining operator $\Omega_{w, w_{12}}^{(i)}$ have the form: 
\begin{enumerate}
\item $e_{w,x}^{(i)}\otimes e_{{w_{12}},x}^{(i)} $ with eigenvalue $0$;
\item ${e}_{w,0}^{(i)}\otimes e_{{w_{12},x}}^{(i)}$ with eigenvalue $x$;
\item $e_{w,x}^{(i)}\otimes{e}_{{w_{12}},0}^{(i)}$ with eigenvalue $-x$;
\item ${e}_{w,0}^{(i)}\otimes {e}_{{w_{12}},0}^{(i)}$ with eigenvalue $0$.
\end{enumerate}
from which we can construct the kernel space for $\bar{\Omega}_{w, w_{12}}^{(i)}$. 
\par 
The singlet space is the intersaction of the two kernel spaces just given.


\section{Orthonormalization and phase conventions}  
The subduction space given by (\ref{subdmbr}) has dimension $\mu$ equal to the multiplicity of $[f, \lambda] \downarrow [f_1, \lambda_1] \otimes [f_2, \lambda_2]$. Then SDCs are not univocally determined. A choice of orthonormality between the different copies of multiplicity imposes a precise form for the multiplicity separations. 
\par
Following the notation given in~\cite{chilla}, let $\{ \chi_1, \ldots, \chi_{\mu} \}$ be a basis in the subduction space. Orthonormality implies for the scalar products:
\begin{equation}
(\chi_{\eta} , \chi_{\eta'}) = \dim(\mathfrak{B}_{f_1}(x), [f_1,\lambda_1]) \dim(\mathfrak{B}_{f_2}(x), [f_2,\lambda_2]) \ \delta_{\eta \eta'}.
\label{onbr} 
\end{equation}
If we denote by $\chi$ the matrix which has the basis vectors of the subduction space as columns, we may orthonormalize it by a suitable $\mu \times \mu $ matrix $\sigma$, i.e. 
\begin{equation}
\tilde{\chi} = \chi \sigma.
\label{ortonbr}   
\end{equation}
In (\ref{ortonbr}) $\tilde{\chi}$ is the matrix which has the orthonormalized basis vectors of the subduction space as columns. Now we can write (\ref{onbr}) as
\begin{equation}
\sigma^t \ \tau \ \sigma = I, 
\label{silvbr}
\end{equation}
where $I$ is the $\mu \times \mu$ identity matrix and $\tau$ is the $\mu \times \mu$ positive defined quadratic form given by
\begin{equation}
\tau = \frac{1}{\dim(\mathfrak{B}_{f_1}(x), [f_1,\lambda_1]) \dim(\mathfrak{B}_{f_2}(x), [f_2,\lambda_2])} \ \chi^{t} \chi.
\end{equation}
From (\ref{silvbr}) we can see $\sigma$ as the Sylvester matrix of $\tau$, i.e. the matrix for the change of basis that reduces $\tau$ in the identity form. We can express $\sigma$ in terms of the orthonormal matrix $O_{\tau}$ that diagonalizes $\tau$
\begin{equation}
\sigma = O_{\tau} D^{-\frac{1}{2}}_\tau O,
\end{equation}
where $D^{-\frac{1}{2}}_{\tau}$ is the diagonal matrix with eigenvalues given by the inverse square root of the eigenvalues of $\tau$ and $O$ a { \it generic} orthogonal matrix. Thus, the general form for the orthonormalized $\chi$ is 
\begin{equation}
\tilde{\chi} = \chi O_{\tau} D^{-\frac{1}{2}}_\tau O. 
\label{genchibr}
\end{equation}
We notice that in case of multiplicity-free subduction, only one choice of global phase has to be made (for example Young-Yamanouchi phase convention~\cite{Chenbook}). It derives from the trivial form of the orthogonal $1 \times 1$ matrices $O$ and $O_{\tau}$.
\par 
To fix the Young-Yamanouchi phase convention we need an ordering relation on permutation lattices (or pair of permutation lattices) and on nodes of the subduction graph. A possible natural choice is the following: given two distinct permutation lattices of order $f$ and diagram $\lambda$, $w=(w_1, w_2, \ldots, w_f)$ and $w'=(w_1', w_2', \ldots, w_f')$, we say that $w < w'$ if the first non-zero element of the word $w-w'=(w_1-w_1', w_2-w_2', \ldots, w_f-w_f')$ is a negative number. Such a relation can be extended to pairs of permutation lattices alphabetically: given two distinct pairs of permutation lattices $w_{12}=(w_1, w_2)$ and $w'_{12}=(w_1', w_2')$, we say that $w_{12}<w_{12}'$ if $w_1 < w_1'$ or $w_1=w_1'$ and $w_2<w_2'$. Resulting from the previous ordering relations, we can provide the ordering relation for nodes of the grid $G=(f;\lambda;f_1, f_2; \lambda_1, \lambda_2)$. For two distinct nodes $n=\langle w; w_{12} \rangle \in G,\ n'=\langle w'; w'_{12} \rangle \in G$ we say $n<n'$ if $w<w'$ or $w=w'$ and $w_{12} < w_{12}'$. 
\par 
Thus, the Young-Yamanouchi phase convention can be stated as follows: we fix to be \emph{positive} the first non-zero SDC with respect to the ordering relation defined on the corresponding nodes.      
\par
We conclude by observing that, in the general case of multiplicity $\mu > 1$, $2^{\mu - 1}$ phases deriving from the $O_{\tau}$ matrix and $1$ phase from the matrix $O$ have to be fixed. Therefore we have $2^{\mu - 1} + 1$ phases to choose. Furthermore we have other $\frac{\mu (\mu -1)}{ 2}$ degrees of freedom deriving from $O$. In sum, as in the case of the symmetric group subduction problem, we have a total of $(2^{\mu - 1} + 1) + \frac{\mu (\mu -1)}{ 2} $ choices to make. 


\section{Outlooks}
There are at least three possible interesting developments for this paper.
\par  
First, one can directly apply the algebraic and combinatoric approach outlined in this paper to Racah-Wigner calculus for \emph{quantized enveloping algebras}. In fact, centralizer algebras (i.e. \emph{Birman-Wenzl} and \emph{type A Iwahori-Hecke algebras}) for quantized enveloping algebras are well characterized both from the algebraic and combinatorial point of view and for the explicit construction of their irreducible representations~\cite{leducram}. Thus, the linear equation method described in terms of subduction graph can be directly applied to this issue without any particular difficulty.  
\par 
Second, Racah-Wigner calculus for \emph{projective representations} of classical Lie groups is very useful in many situations. For example it is often necessary when one has to describe the states of physical systems involving fermions. Finding such representations is equivalent to determine the tensorial irreducible representations of the \emph{universal enveloping group} of the original Lie group $G$. An alternative approach is to find the projective representations of Brauer algebras. The Gelfand-Tzetlin basis for such representations is described in terms of combinatorial objects which are known as \emph{stable-up-down tableaux}~\cite{koike} (they are \emph{permutation lattices with null elements}, in the language of this paper). Unfortunately, the explicit action on the irreducible modules are still unknown.
\par       
Finally, the Racah-Wigner calculus for exceptional Lie groups also have many application both in physics and mathematical physics. The study of Racah would be important for a comprehensive knowledge of the Racah-Wigner calculus for all Lie groups.


\end{document}